\documentclass[aps,prb,twocolumn,floatfix,showpacs]{revtex4}

\usepackage{graphicx}% Include figure files
\usepackage{dcolumn}% Align table columns on decimal point
\usepackage{bm}% bold math
\usepackage{verbatim}

\newcommand{\STR}[3]{{#1}({#2}$\times${#3})}
\newcommand{\III}{\ensuremath{\mathrm{III}}}
\newcommand{\V}{\ensuremath{\mathrm{V}}}

\begin{document}

\title{Structure of AlSb(001) and GaSb(001) Surfaces Under Extreme
  Sb-rich Conditions}

\author{Jeffery~Houze} 
\author{Sungho~Kim} 
\author{Seong-Gon~Kim} \email{kimsg@ccs.msstate.edu} 
\affiliation{
  Department of Physics and Astronomy, 
  Mississippi State University,
  Mississippi State, MS 39762, USA
}
\affiliation{%
  Center for Advanced Vehicular Systems, 
  Mississippi State University, 
  Mississippi State, MS 39762, USA
}

\author{S.~C.~Erwin}
\author{L.~J.~Whitman}
\affiliation{Naval Research Laboratory, Washington, DC 20375, USA}

\date{June 15, 2007}

\begin{abstract}
  
  We use density-functional theory to study the structure of AlSb(001)
  and GaSb(001) surfaces.  Based on a variety of reconstruction
  models, we construct surface stability diagrams for AlSb and GaSb
  under different growth conditions.  For AlSb(001), the predictions
  are in excellent agreement with experimentally observed
  reconstructions. For GaSb(001), we show that previously proposed
  model accounts for the experimentally observed reconstructions under
  Ga-rich growth conditions, but fails to explain the experimental
  observations under Sb-rich conditions.  We propose a new model that
  has a substantially lower surface energy than all
  \STR{}{$n$}{5}-like reconstructions proposed previously and that, in
  addition, leads to a simulated STM image in better agreement with
  experiment than existing models.  However, this new model has higher
  surface energy than some of \STR{}{4}{3}-like reconstructions,
  models with periodicity that has not been observed. Hence we
  conclude that the experimentally observed \STR{}{1}{5} and
  \STR{}{2}{5} structures on GaSb(001) are kinetically limited rather
  than at the ground state.

\end{abstract}

\pacs{%
68.35.Bs, % Structure of clean surfaces (reconstruction) 
68.35.Md, % Surface thermodynamics, surface energies
68.37.Ef, % Scanning tunneling microscopy
          % (including chemistry induced with STM) 
68.47.Fg  % Semiconductor surfaces 
}

\maketitle

\section{Introduction}

The surfaces and interfaces of III-V semiconductors constitute some of
the most important components of the semiconductor industry.  For
example, III-V heterostructure quantum wells are key components in a
wide range of optical and high-frequency electronic devices, including
field-effect transistors \cite{Boos:1998}, resonant tunneling
structures \cite{Scott:1994}, infrared lasers \cite{Yang:1998}, and
infrared detectors \cite{Fuchs:1997}. Many of these devices require
extremely sharp and clean interfaces. For this reason, an
understanding of the atomic-scale morphology of III-V semiconductor
surfaces is critical to controlling the growth and formation of their
interfaces \cite{Klimeck:1998, Ting:1996}.
 
It is generally accepted that the surfaces of III-V semiconductors
should reconstruct in such a way that the number of electrons is
exactly enough to doubly occupy all orbitals on electronegative (V)
atoms, leaving all orbitals on electropositive (III) atoms unoccupied.
This guiding principle, known as the electron-counting model (ECM),
has been used to screen candidate structural models of many observed
reconstructions on the surfaces of III-V semiconductors
\cite{Chadi:1987, Pashley:1989, Northrup:1994, Zhang:1996,
  MacPherson:1996}.  In practice, however, not all experimentally
realized reconstructions follow this principle. For example, under
Sb-rich growth conditions, GaSb(001) forms surface reconstructions
that are weakly metallic and hence violate the ECM
\cite{Whitman:1997}, even though the closely related AlSb(001) surface
forms insulating reconstructions that satisfy it \cite{Barvosa:2000}.
The nature of reconstructions that violate the ECM, and the underlying
reasons for their stability, are thus important for understanding
III-V surfaces in general.

In this article we explore theoretically a large number of judiciously
chosen candidate reconstructions on GaSb(001) and AlSb(001).  We find
that as the growth conditions are varied between Sb-poor and Sb-rich,
the predicted sequence of stable reconstructions for GaSb(001) is
exactly analogous to those of AlSb(001).  Experimentally, however, the
picture is more complicated.  In the Sb-poor limit, the observed
GaSb(001) reconstruction is indeed analogous to that of AlSb(001).  On
the other hand, in the Sb-rich limit, the experimentally observed
reconstructions for GaSb(001) and AlSb(001) are different.  Moreover,
in this limit the predicted and observed reconstructions are in good
agreement only for AlSb(001), while for GaSb(001) there remains an
unresolved discrepancy between theory and experiment.

Experimentally, the Sb-terminated AlSb(001) surface evolves through
the sequence \STR{$\alpha$}{4}{3} $\rightarrow$ \STR{$\beta$}{4}{3}
$\rightarrow$ \STR{$\gamma$}{4}{3} $\rightarrow$ \STR{c}{4}{4} as the
growth condition is changed from low Sb flux (or high substrate
temperature) to high Sb flux (or low temperature) \cite{Barvosa:2000}.
All of these reconstructions are insulating, and are well accounted
for by structural models proposed in the literature that satisfy the
ECM. 

Of particular interest here is the Sb-rich AlSb(001)-\STR{$c$}{4}{4}
reconstruction, analogous to the As-rich GaAs(001)-\STR{$c$}{4}{4}
reconstruction, which is observed on AlSb but not GaSb. In contrast to
both AlSb and GaAs, the GaSb(001) surface does not exhibit a stable,
insulating \STR{$c$}{4}{4} reconstruction under similar---or any
other---growth conditions.  Instead, it forms \STR{}{$n$}{5}-like
reconstructions \cite{Sieger:1995, Piao:1990, Yano:1991,
  Franklin:1990, Whitman:1997}.  Structural models proposed in the
literature for these \STR{}{$n$}{5}-like reconstructions violate, by
design, the ECM and consequently are weakly metallic
\cite{Whitman:1997}.  Simulated scanning tunneling microscopy (STM)
images based on \STR{}{2}{10} and \STR{$c$}{2}{10} models closely
resemble the experimental images \cite{Whitman:1997}. As a result,
these models have been generally accepted as describing the GaSb(001)
surface under Sb-rich growth conditions.  Nevertheless, we show below
on energetic grounds that these models are unlikely to be correct.
Specifically, we find their calculated surface energy to be
significantly higher than GaSb(001)-\STR{$c$}{4}{4} for any plausible
value of Sb chemical potential.  However, since the experimentally
observed reconstruction of GaSb(001) does not have \STR{$c$}{4}{4}
periodicity, this model cannot be correct either.  Thus a definitive
structural model remains to be found.

\section{Methods}

The basic structural models we considered are taken from the literature and
are shown in Figs.~\ref{fig:models-1} and \ref{fig:models-2}.  Surfaces
that satisfy the ECM are generally semiconducting, while those that do
not may be metallic.  The degree to which a given surface satisfies
the ECM can be measured by the excess electron count, $\Delta\nu$,
which we define here as the difference between the number of available
electrons and the number required to satisfy the ECM, per \STR{}{1}{1}
surface unit cell.  Excess electron counts for the structural models
in Figs.~\ref{fig:models-1} and \ref{fig:models-2} are tabulated in
Table~\ref{tab:ECM}.

To compare the surface energies of reconstruction models with
different periodicities and stoichiometries, we consider the surface
energy per unit area,
\begin{equation}
\gamma  =  E_{\text{surf}}/A  = (E_{\text{tot}} 
- n_{\III}\mu'_{\III} - n_{\V}\mu'_{\V})/A,
\label{eq:E-surf-1}
\end{equation}
where $E_{\text{tot}}$ is the total energy of a reconstructed surface,
of area $A$, containing $n_{\III}$ group-III and $n_{\V}$ group-V
adatoms in excess with respect to the bulk-truncated, Sb-terminated
AlSb(001) or GaSb(001).  The atomic chemical potentials $\mu'$ are
more conveniently expressed in terms of excess chemical potentials
$\mu$, relative to the energy per atom in the ground-state elemental
phases: $\mu' = \mu^{\text{bulk}}+\mu$.  Assuming the surface to be in
thermodynamic equilibrium with the bulk, the III and V chemical
potentials are related by $\mu_{\III}$ + $\mu_{\V}$ = $\Delta H_f$,
where $\Delta H_f = \mu_{\text{III-V}}^{\text{bulk}} -
(\mu_{\III}^{\text{bulk}} + \mu_{\V}^{\text{bulk}})$ is the formation
enthalpy of the bulk III-V crystalline phase \cite{Qian:1988} (note
that $\Delta H_f$ is intrinsically negative).  Eq.~(\ref{eq:E-surf-1})
can then be rewritten to show more clearly the dependence of $\gamma$
on the surface stoichiometry and chemical potential:
\begin{equation}
\gamma =  \gamma_0 + \mu_{\V} \Delta\Theta.
\label{eq:E-surf-2}
\end{equation}
Here $\gamma_0 = (E_t - E_{\text{sub}}) -
\mu_{\text{III-V}}^{\text{bulk}}\Theta_{\III} +
\mu_{\V}^{\text{bulk}}\Delta\Theta$ is independent of the chemical
potentials, and $\Delta\Theta = \Theta_{\III} - \Theta_{\V} =
(n_{\III} - n_{\V})/A$ is the deviation of the surface stoichiometry
from its bulk value.  The dependence of $\gamma$ on chemical potential
is given entirely by the second term.  Note that $\mu_{\V}$ is
intrinsically negative, and can take values in the range $\Delta H_f
\le \mu_{\V} \le 0$.  Hence, Eq.~(\ref{eq:E-surf-2}) reflects in a
simple way that III-rich reconstructions ($\Theta_{\III} >
\Theta_{\V}$) are favored under III-rich conditions ($\mu_{\V}
\rightarrow \Delta H_f$), V-rich reconstructions ($\Theta_{\V} >
\Theta_{\III}$) are favored under V-rich conditions ($\mu_{\V}
\rightarrow 0$), and for stoichiometric reconstructions ($\Theta_{\V}
= \Theta_{\III}$) $\gamma$ does not depend on chemical potential.

To compute the total-energy contribution, $\gamma_0$, to the surface
energy we performed first-principles calculations using
density-functional theory (DFT).  The calculations were performed
within the local-density approximation (LDA) \cite{Ceperley:1980,
  Perdew:1981} using ultrasoft pseudopotentials \cite{Vanderbilt:1990,
  Kresse:1994, Kresse:1996}. We used a standard supercell technique,
modeling the (001) surface with a slab consisting of four bilayers
and 10 \AA\ of vacuum. Atoms in the bottom bilayer were fixed
at their bulk positions, while all other atoms are allowed to relax
until the rms force was less than 0.005~eV/\AA. The bottom layer
(either Ga or Al) was passivated with pseudohydrogens. A
plane-wave cutoff of 300~eV was used, and reciprocal space was sampled
with a density equivalent to at least 192 $k$-points in the
\STR{}{1}{1} surface Brillouin zone.  To define the III-V formation
enthalpy $\Delta H_f$ from the bulk chemical potentials
$\mu^{\text{bulk}}$, separate DFT calculations were performed for the
elements in their ground-state phases: Ga in the $\alpha$-Ga
structure, Al in the face-centered cubic structure, Sb in the
rhombohedral structure, and both AlSb and GaSb in the zinc blende
structure.

\section{Results and Discussions}

\begin{table*}
  \caption{\label{tab:ECM} Electron count for different reconstructions
    of GaSb(001) surface. The excess electron count per \STR{}{1}{1}
    surface unit cell is defined as $\Delta\nu = (\tilde{n} - \tilde{m})/A$
    where $\tilde{n}$ is the number of available electrons and
    $\tilde{m}$ is the number of required electrons to satisfy the ECM
    in the excess of Sb-terminated GaSb(001). $A$ is the area of the 
    surface unit cell in terms of the \STR{}{1}{1} surface unit cell. 
    $n_i$ is number of adatoms of species $i$ in excess with respect
    to the Sb-terminated GaSb(001) and $\Theta_i = n_i/A$ is the 
    coverage of adatoms of species $i$.
    The relative $\gamma$ values, in eV per \STR{}{1}{1} surface unit 
    cell, are given with respect to that of \STR{$\alpha$}{4}{3}.}
  \begin{ruledtabular}
    \begin{tabular}{cccccccccccc}
      Structure & $A$ & $n_{\III}$ & $n_{\V}$ 
      & $\Theta_{\III}$ & $\Theta_{\V}$ & $\Delta\Theta$
      & $\tilde{n}$ & $\tilde{m}$ & $\Delta\nu$ 
      & $\gamma$(Ga-rich) & $\gamma$(Sb-rich) \\
      \hline
      \STR{$\alpha$}{4}{3} & 12 & 4 & 4 & 0.333 & 0.333 & 0.0
      & 62 & 62 & 0 & 0.000 & 0.000 \\
      \STR{$\beta$}{4}{3} & 12 & 1 & 7 & 0.083 & 0.583 & -0.5
      & 68 & 68 & 0 & 0.076 & -0.074 \\
      \STR{$\gamma$}{4}{3} & 12 & 0 & 8 & 0.0 & 0.667 & -0.667
      & 70 & 70 & 0 & 0.114 &  \textbf{-0.087} \\
      \STR{$h0$}{4}{3} & 12 & 0 & 8 & 0.0 & 0.667 & -0.667
      & 70 & 70 & 0 & 0.118 & -0.083 \\
      \STR{$c$}{4}{4} & 8 & 0 & 6 & 0.0 & 0.750 & -0.750
      & 50 & 50 & 0 & 0.142 & -0.084 \\
      \hline
      \STR{$c$}{2}{10} & 10 & 0 & 8 & 0.0 & 0.800 & -0.800
      & 65 & 62 & 0.3 & 0.255 & 0.014 \\
      \STR{}{2}{10} & 20 & 0 & 24 & 0.0 & 1.200 & -1.200
      & 170 & 164 & 0.3 & 0.528 & 0.166 \\
      \hline
      \STR{s1a-$c$}{2}{10} & 10 & 1 & 7 & 0.1 & 0.700 & -0.600 
      & 63 & 62 & 0.1 & 0.143 & \textbf{-0.038} \\
      \STR{s1b-$c$}{2}{10} & 10 & 1 & 7 & 0.1 & 0.700 & -0.600
      & 63 & 62 & 0.1 & 0.181 & 0.000 \\
      \STR{s1c-$c$}{2}{10} & 10 & 1 & 7 & 0.1 & 0.700 & -0.600
      & 63 & 62 & 0.1 & 0.280 & 0.099 \\
      \hline
      \STR{s2a-$c$}{2}{10} & 10 & 2 & 6 & 0.2 & 0.600 & -0.400
      & 61 & 62 & -0.1 & 0.137 & 0.016 \\
      \STR{s2b-$c$}{2}{10} & 10 & 2 & 6 & 0.2 & 0.600 & -0.400
      & 61 & 62 & -0.1 & 0.124 & 0.003 \\
      \STR{s2c-$c$}{2}{10} & 10 & 2 & 6 & 0.2 & 0.600 & -0.400
      & 61 & 62 & -0.1 & 0.143 & 0.023 \\
      \STR{s2d-$c$}{2}{10} & 10 & 2 & 6 & 0.2 & 0.600 & -0.400
      & 61 & 62 & -0.1 & 0.141 & 0.020 \\
      \STR{s2e-$c$}{2}{10} & 10 & 2 & 6 & 0.2 & 0.600 & -0.400
      & 61 & 62 & -0.1 & 0.167 & 0.046 \\
      \STR{s2f-$c$}{2}{10} & 10 & 2 & 6 & 0.2 & 0.600 & -0.400
      & 61 & 62 & -0.1 & 0.144 & 0.023 \\
      \STR{s2g-$c$}{2}{10} & 10 & 2 & 6 & 0.2 & 0.600 & -0.400
      & 61 & 62 & -0.1 & 0.182 & 0.062 \\
      \STR{s2h-$c$}{2}{10} & 10 & 2 & 6 & 0.2 & 0.600 & -0.400
      & 61 & 62 & -0.1 & 0.164 & 0.043 \\
      \STR{s2i-$c$}{2}{10} & 10 & 2 & 6 & 0.2 & 0.600 & -0.400
      & 61 & 62 & -0.1 & 0.166 & 0.045 \\
     \end{tabular}
  \end{ruledtabular}
\end{table*}

\begin{figure}
  \resizebox{\hsize}{!}{\includegraphics{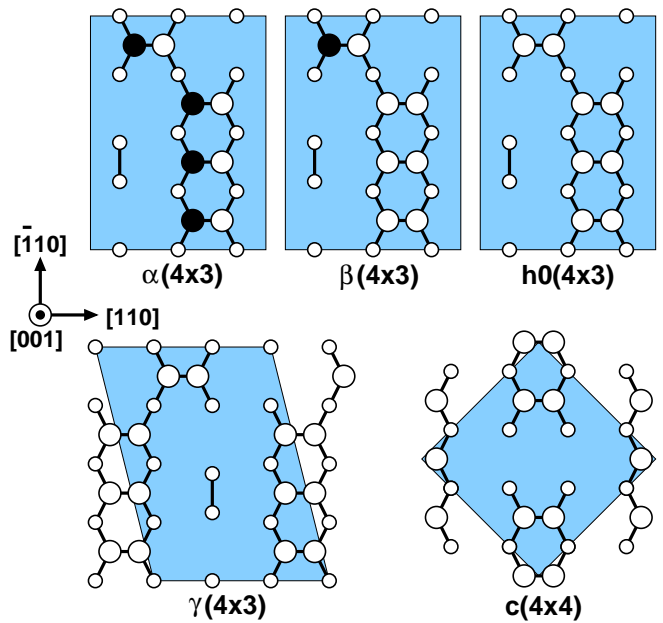}}
  \caption{\label{fig:models-1} (color online) Reconstruction models
    proposed for the AlSb(001) or GaSb(001) surfaces with \STR{}{4}{3}
    and \STR{}{4}{4} periodicities. The first two upper layers are
    shown in a top view.  Smaller white circles represent Sb atoms in
    the top layer of the underlying Sb-terminated AlSb(001) or
    GaSb(001) surface.  Larger circles represent Al or Ga (black) and
    Sb (white) adatoms.  The unit cells are shown in light blue.}
\end{figure}

\begin{figure}
  \resizebox{\hsize}{!}{\includegraphics{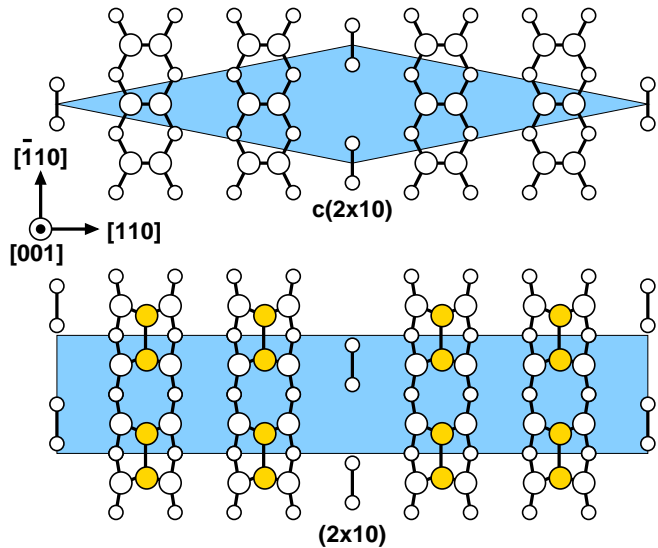}}
  \caption{\label{fig:models-2} (color online) Reconstruction models
    proposed for the GaSb(001)-\STR{}{1}{5}-like surfaces under the
    extreme Sb-rich growth condition. See Fig.~\ref{fig:models-1} for
    color schemes.  Gold circles represent the second layer Sb
    adatoms. }
\end{figure}

\begin{figure}
  \resizebox{\hsize}{!}{\includegraphics{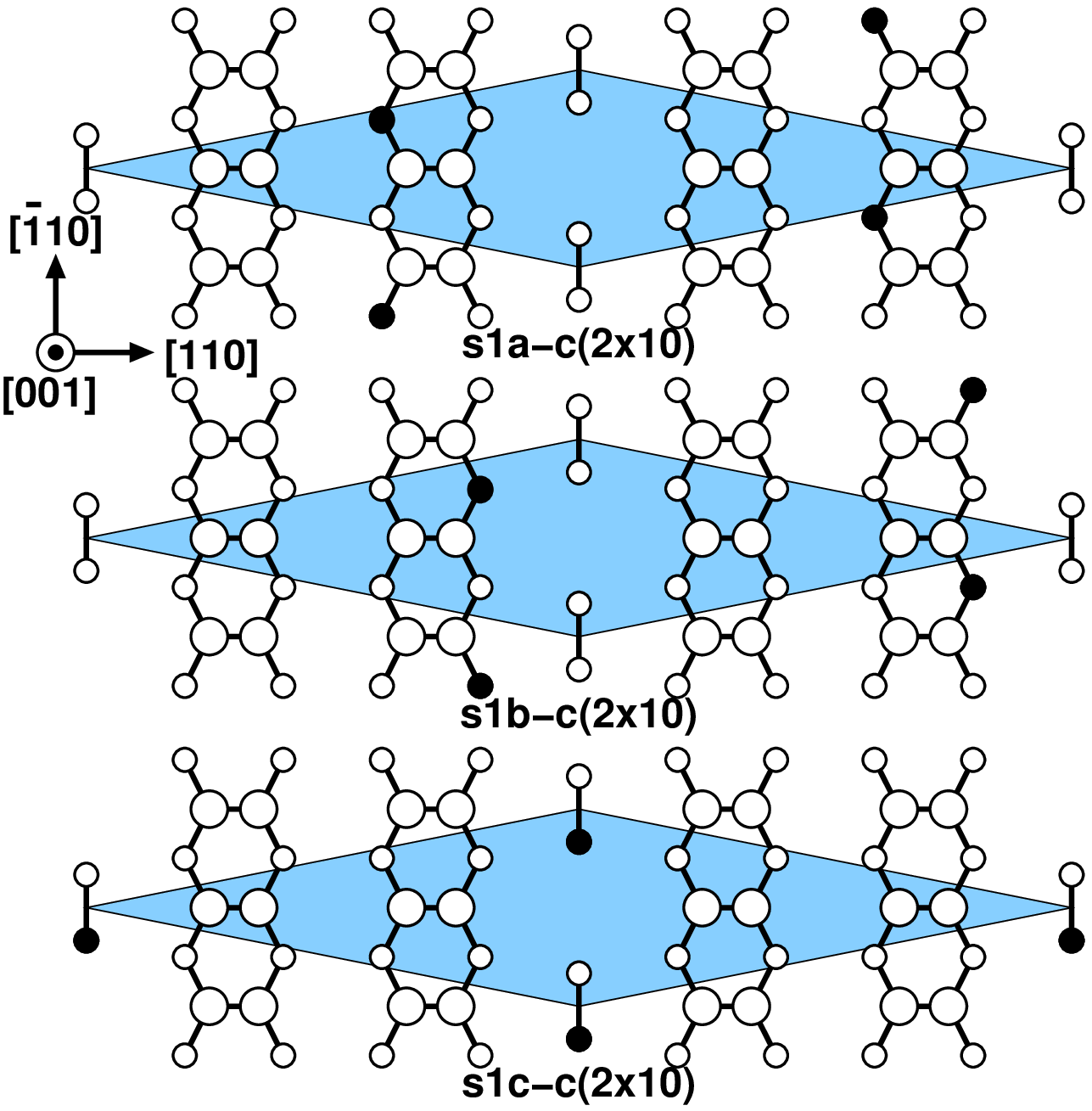}}
  \caption{\label{fig:models-s1} (color online) Reconstruction models
    with a single substitution of Sb atoms by Ga atoms.  See
    Fig.~\ref{fig:models-1} for color schemes.  }
\end{figure}

\begin{figure*}
  \resizebox{\hsize}{!}{\includegraphics{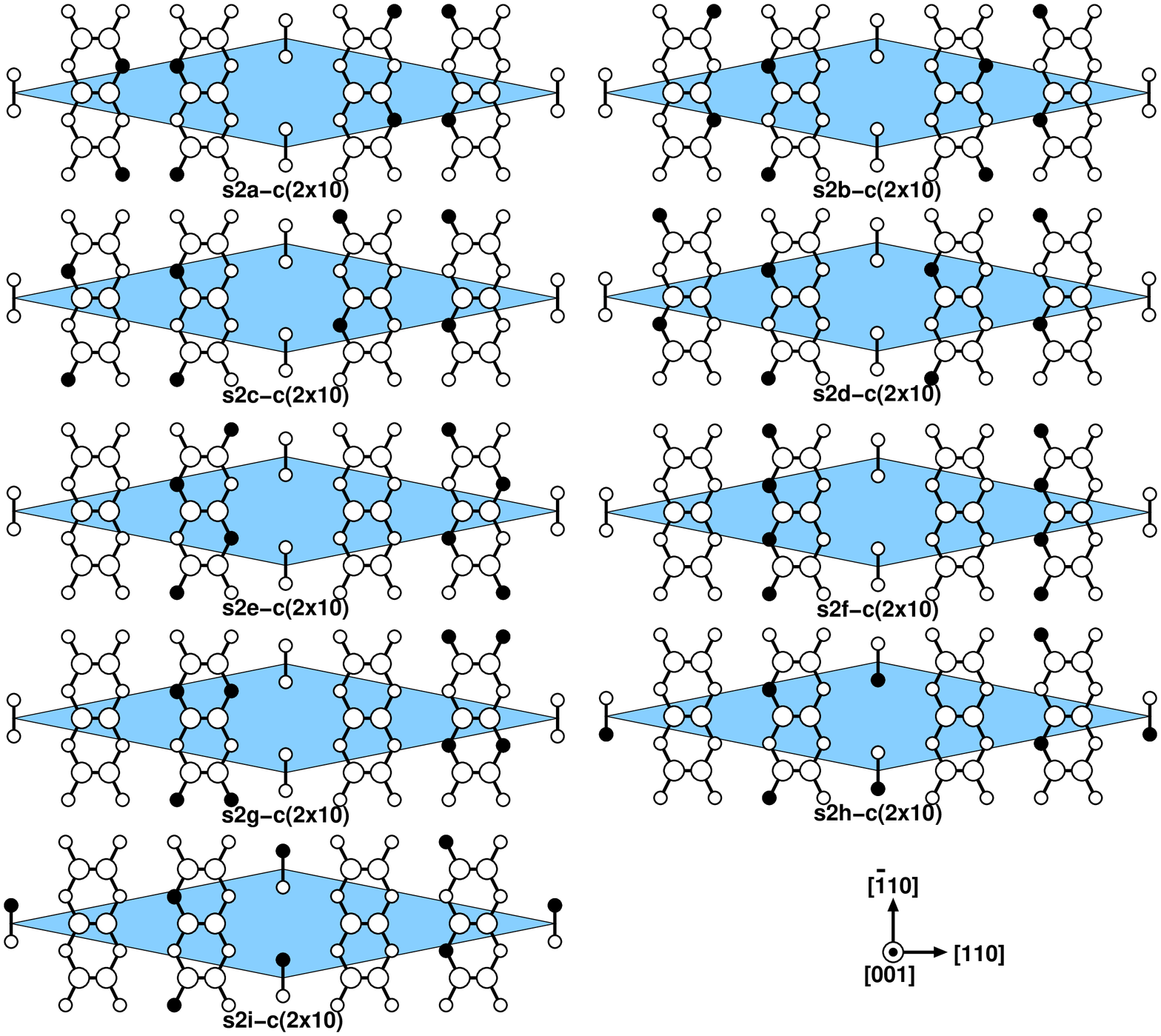}}
  \caption{\label{fig:models-s2} (color online) Reconstruction models
    with a double substitution of Sb atoms by Ga atoms.  See
    Fig.~\ref{fig:models-1} for color schemes.  }
\end{figure*}

\begin{figure}
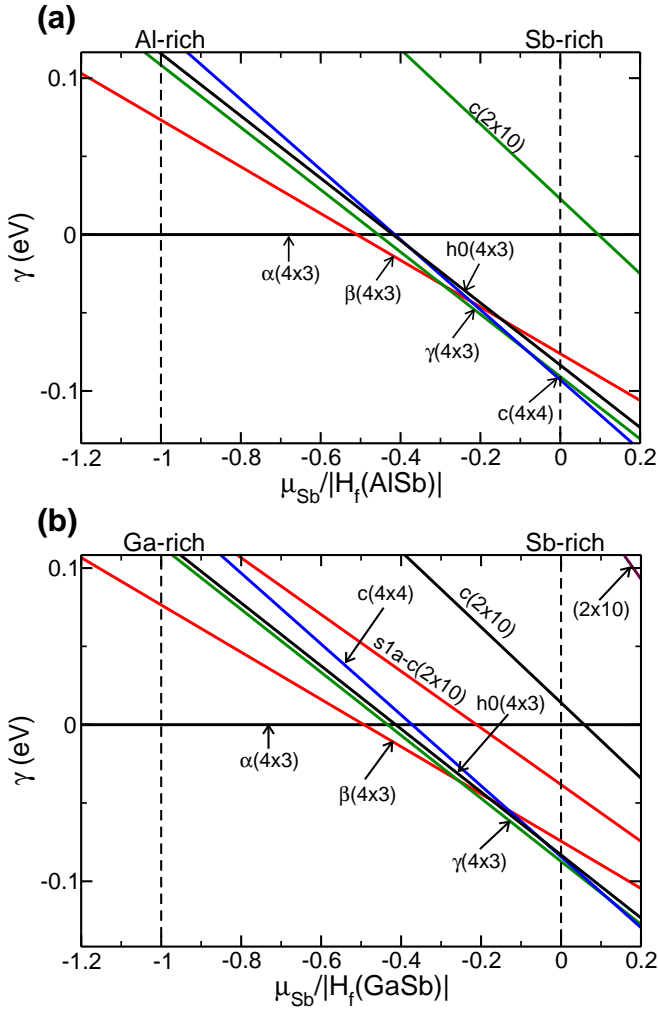

  \resizebox{\hsize}{!}{\includegraphics{Eform-AlSb.eps}}
  \resizebox{\hsize}{!}{\includegraphics{Eform-GaSb.eps}}
  \caption{\label{fig:surf-stab} (color online) (a) Surface stability
    phase diagram for AlSb(001) surface. The relative surface energy
    [Eq.~(\ref{eq:E-surf-2})] is plotted as a function of the Sb
    chemical potential relative to its corresponding bulk
    value. Dotted vertical lines mark the thermodynamically allowed
    range of $\mu_{\text{Sb}}$. $\Delta H_f$ is the heat of formation
    for AlSb. (b) Surface stability phase diagram for GaSb(001)
    surface; $\Delta H_f$ is the heat of formation for GaSb.}
\end{figure}

The resulting relative surface energies for AlSb(001) and GaSb(001)
are shown in Figs.~\ref{fig:surf-stab}(a) and \ref{fig:surf-stab}(b),
respectively, for the eight models considered here.  For each model,
the surface energy is linear in $\mu_{\V}$, with the slope given by
$\Delta\Theta$.

For AlSb(001) the predicted stable reconstructions, and their
energetic ordering, are in excellent agreement with experiment.
Proceeding from Sb-poor to Sb-rich conditions, the predicted sequence
is \STR{$\alpha$}{4}{3} $\rightarrow$ \STR{$\beta$}{4}{3}
$\rightarrow$ \STR{$\gamma$}{4}{3} $\rightarrow$ \STR{$c$}{4}{4}, as
reported previously \cite{Barvosa:2000}.  This is the same sequence
observed experimentally \cite{Barvosa:2000}.  Moreover,
\STR{$\gamma$}{4}{3} is predicted to exist only over a very narrow
range of $\mu_{\text{Sb}}$, in agreement with experiment
\cite{Barvosa:2000}.

For GaSb(001) the predicted sequence is qualitatively the same as for
AlSb(001), although the \STR{$c$}{4}{4} is only predicted to be stable
for values of $\mu_{\text{Sb}}$ above the thermodynamically allowed
limit of zero.  Experimentally, however, the situation is quite
different.  As reported previously, neither the \STR{$\gamma$}{4}{3}
nor the \STR{$c$}{4}{4} phase is observed for any growth condition
\cite{Barvosa:2000}. Instead, under Sb-rich conditions, only the
\STR{}{1}{5} and \STR{}{2}{5} periodicities have been observed
\cite{Whitman:1997}.

\begin{figure}
  \resizebox{\hsize}{!}{\includegraphics{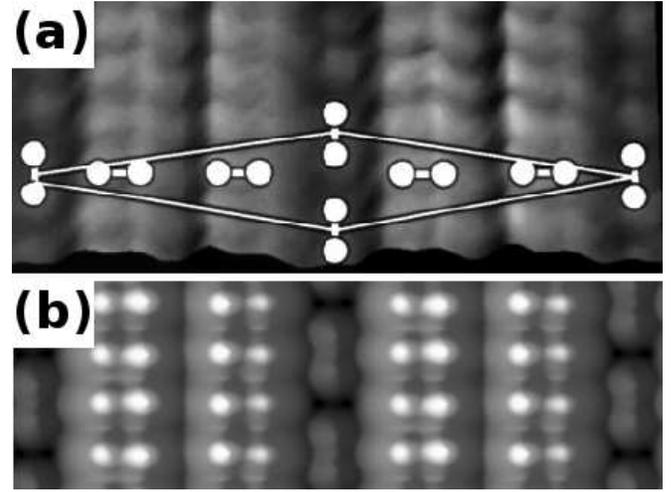}}
  \caption{\label{fig:STM} Filled-state STM images of GaSb(001) with
    \STR{}{1}{5} periodicity.  (a) Experimental STM image; a
    \STR{$c$}{2}{10} unit cell is indicated.  (b) Simulated STM image
    of \STR{s1a-$c$}{2}{10}.  This image shows the asymmetries in the
    intensities of the current density from two atoms of the
    horizontal dimers, which was not captured in the simulated STM
    image of \STR{$c$}{2}{10}.  Compare with Fig.~3(d) of
    Ref.~\onlinecite{Whitman:1997}.}
\end{figure}
Righi \textit{et al.} suggested \STR{$h0$}{4}{3}, shown in
Fig.~\ref{fig:models-1}, as the model for GaSb(001) surface under these
conditions \cite{Righi:2005}. Our calculation indeed shows that it is
energetically as favorable as \STR{$\gamma$}{4}{3}, as shown in
Fig.~\ref{fig:surf-stab}(b) and in Table~\ref{tab:ECM}. However,
\STR{$h0$}{4}{3} must be rejected as it has a wrong periodicity.  

In order to explain the experimentally observed \STR{}{1}{5} and
\STR{}{2}{5} structures on GaSb(001) surface, we studied a large
number of structures based on variations of \STR{$c$}{2}{10} and
\STR{}{2}{10}.  We note that \STR{$c$}{2}{10} violates the ECM
substantially ($\Delta\nu = 0.3$) and substitution of Sb atoms in the
top layer of the underlying Sb-terminated GaSb(001) surface by Ga
atoms can lower the excess electron count.  Fig.~\ref{fig:models-s1}
shows the possible reconstructions when a single Sb atom is replaced
by a Ga atom.  We use the naming convention of $s1x$ to denote a
``single substitution''.  As shown in Table~\ref{tab:ECM}, all $s1x$
reconstruction indeed have lower excess electron counts.

For completeness, we also considered reconstructions resulting from
double substitution of Sb atoms by Ga atoms as shown in
Fig.~\ref{fig:models-s2}. More substitutions, however, were not found
to be energetically favorable: Table~\ref{tab:ECM} shows that the
surface energy of these structures are higher than that of $s1x$
reconstructions.  We note that for these double substitutions the
excess electron counts $\Delta\nu$ are negative, indicating a deficit
of electrons relative to the ECM.

One of the most energetically favorable structures having the correct
periodicity is \STR{s1a-$c$}{2}{10}, shown in
Fig.~\ref{fig:models-s1}.  \STR{s1a-$c$}{2}{10} has two clear
advantages over \STR{$c$}{2}{10}.  First, the surface energy
for \STR{s1a-$c$}{2}{10} is lower than that of
\STR{$c$}{2}{10} by more than 50 meV per \STR{}{1}{1} unit
cell. Second, as shown in Fig.~\ref{fig:STM}, the simulated STM image
for \STR{s1a-$c$}{2}{10} is in a better agreement with the experiment
image, in that it reproduces the left-right asymmetry within the
surface Sb dimers \cite{Whitman:1997}.  Furthermore, as shown in
Table~\ref{tab:ECM}, this model violates the ECM and thus is predicted
to be weakly metallic, as observed in tunneling spectroscopy
\cite{Feenstra:1994}.  Therefore, the previously proposed model
\STR{$c$}{2}{10} is unlikely to be the experimentally
realized structure.  

However, the calculated surface energy of \STR{s1a-$c$}{2}{10} is
{\em higher} than that of \STR{$\gamma$}{4}{3}, as shown in
Table~\ref{tab:ECM} and Fig.~\ref{fig:surf-stab}(b).  Likewise,
\STR{}{2}{10}, the structural model generally accepted for the surface
with \STR{}{2}{5} periodicity, is the least energetically favorable
structure among the eight structures of Table~\ref{tab:ECM}.  On the
other hand, \STR{$\gamma$}{4}{3}, the most energetically favorable
structure among all the structures considered in this study, has a
periodicity that has not been observed experimentally to date.  These
facts leave us with two possible conclusions: either the correct
structural model remains undiscovered, or the experimentally obtained
surface is not the ground-state structure.

The latter possibility, a kinetically limited surface, bears closer
consideration.  For example, there may be an activation barrier to
forming the mixed dimers on GaSb that cannot be overcome within the
growth temperatures and times used here.  Indeed, to stabilize these
surfaces during the growth, one must go from active growth with both
Ga and Sb flux at $\sim$500~$^\circ$C, to room temperature and no flux
while trying to stabilize the surface.  This process typically
involved simultaneously lowering the temperature while turning off the
Ga and then lowering the Sb flux.  The surface cannot be annealed,
because that would drive off Sb and create \STR{}{$n$}{3}
reconstructions.  These considerations lead us to propose that the
\STR{s1a-$c$}{2}{10} structure is the most likely model for the
observed GaSb(001) surface as created under Sb-rich growth conditions
and subsequently stabilized under vacuum.

\section{Summary and Conclusions}

We have performed \textit{ab initio} calculations on the surface
energy and atomic structure of AlSb(001) and GaSb(001) surfaces with
various reconstructions. Surface stability diagrams for a large number
of reconstruction models are constructed under different growth
conditions.  For AlSb(001), we confirmed that the predictions of the
currently accepted models are in good agreement with experimentally
observed reconstructions.  For GaSb(001), we showed that previously
proposed model accounts for the experimentally observed
reconstructions under Ga-rich growth conditions, but fails to explain
the experimental observations under Sb-rich conditions.  Therefore, we
propose \STR{s1a-$c$}{2}{10} as a better alternative to existing
models for GaSb(001) under extreme Sb-rich growth conditions.  Our
calculations show that \STR{s1a-$c$}{2}{10} has a substantially
lower surface energy than all \STR{}{$n$}{5}-like reconstructions
proposed previously and, in addition, it leads to a simulated STM
image in better agreement with experiment than existing models.
However, \STR{s1a-$c$}{2}{10} has higher surface energy than
\STR{$\gamma$}{4}{3}, a model with periodicity that has not been
observed. Hence we
conclude that the experimentally observed \STR{}{1}{5} and
\STR{}{2}{5} structures on GaSb(001) are not the ground-state
structure, but kinetically limited ones.

\section{Acknowledgment}

This work was in part supported by the US Department of Defense
under the CHSSI MBD-04 (Molecular Packing Software for \textit{ab
  initio} Crystal Structure and Density Predictions) project and by
the Office of Naval Research.  Computer time allocation has been
provided by the High Performance Computing Collaboratory (HPC$^2$) at
Mississippi State University.

\bibliography{DFT,surface}

\end{document}